\documentclass[iop]{emulateapj} 

\usepackage{graphics} 

\def\pb{Pa$\beta$}
\def\bg{Br$\gamma$}
\def\feii{[Fe\,{\sc ii}]}
\def\pii{[P\,{\sc ii}]}

\def\h2{H$_2$}
\def\kms {$\rm km\,s^{-1}$}

\shorttitle{Outflow in NGC\,5929}
\shortauthors{Riffel, Storchi-Bergmann \& Riffel}

\begin{document}




\title{An outflow perpendicular to the radio jet in the Seyfert nucleus of NGC\,5929}


\author{Rogemar A. Riffel
} \affil{Universidade Federal de Santa Maria, Departamento de F\'\i sica, CCNE, 
 97105-900, Santa Maria, RS, Brazil} \email{rogemar@ufsm.br}
\author{Thaisa Storchi-Bergmann}\email{thaisa@ufrgs.br}
\author{Rog\'erio Riffel
}
\affil{Universidade Federal do Rio Grande do Sul, IF, CP 15051, Porto
Alegre 91501-970, RS, Brazil} \email{riffel@ufrgs.br}

\begin{abstract} 
We report the observation of an outflow perpendicular to the radio jet in near-infrared integral field spectra
of the inner 250~pc of the Seyfert\,2 galaxy NGC\,5929. The observations were obtained with the Gemini Near infrared Integral Field Spectrograph at a spatial resolution of $\sim$20\,pc and spectral resolution R$\approx$\,5300 and reveal a region $\sim\,50$\,pc wide crossing the nucleus and extending by $\sim300$\,pc  perpendicularly to the known radio jet in this galaxy. Along this structure -- which we call SE-NW strip -- the emission-line profiles show two velocity components, one blueshifted and the other redshifted by $-150$\kms\ and 150\kms, respectively, relative to the systemic velocity. We interpret these two components as due to an outflow  perpendicular to the radio jet, what is supported by low frequency radio emission observed along the same region. We attribute this feature to the interaction of ambient gas with an ``equatorial outflow''  predicted in recent accretion disk and torus wind models. Perpendicularly to the SE-NW strip, thus approximately along the radio jet, single component profiles show blueshifts of  $\approx-150$\kms\ to the north-east and similar redshifts to the south-west, which can be attributed to gas counter-rotating relative to the stellar kinematics.
 More double-peaked profiles are observed in association with the two radio 
hot-spots, attributed to interaction of the radio jet with surrounding gas.

\end{abstract}





\section{Introduction}

Signatures of interactions between radio jets and the narrow line region (NLR) emitting gas in Seyfert galaxies have been frequently reported in the literature and usually appear in the data as distorted emission line profiles relative to a Gaussian. Theses profiles show, for example, increased velocity dispersions, red or blue wings and other kinematic signatures attributed to outflows observed in regions co-spatial with radio structures \citep[e.g.][]{falcke98, cecil02, whittle04,eso428,mrk1066-kin}. Fast shocks produced by the radio jet can modify the shape of the NLR by compressing the gas and producing line emission enhancement in regions next to radio structures. Usually, the interaction of the radio jet with the NLR is seen only in association with the high frequency radio emission (e.g., from 1.6 to 15 GHz). On the other hand, outflows observed in Seyfert galaxies are not usually originated by the interaction of the radio jet with the NLR gas, but by the interaction of winds from the accretion disk with gas from the circumnuclear region of the galaxy. These outflows have been observed to have a bi-conical shape in a number of nearby Seyfert galaxies \citep[e.g.][]{sb92,das05,das07,sb10}. 

NGC\,5929 is a spiral galaxy harboring a Seyfert\,2 nucleus, located at a distance 35.9\,Mpc\footnote{as quoted in NASA/IPAC EXTRAGALACTIC DATABASE --http://ned.ipac.caltech.edu/}, for which 1 arcsec corresponds to  $\sim$175\,pc at the galaxy. It presents a well defined bi-polar radio jet oriented along the position angle PA$\approx$60$^\circ$, showing a triple structure with two bright hot spots, one located at 0\farcs5 north-east from the nucleus and the other at 0\farcs6 south-west from it. The third and fainter radio structure is observed at the nucleus of the galaxy \citep{ulvestad84,wilson89,su96}. Besides this main jet, weak extended emission is observed at 400\,MHz, perpendicularly to the main radio jet \citep{su96}, and approximately along the minor axis of the galaxy. NGC\,5929 seems to be in interaction with its companion galaxy NGC\,5930.

The gas kinematics of the nuclear region of NGC\,5929 has been discussed in several previous works, but its origin is still not well understood \citep{keel85,ferruit99,rosario10}. 

In this letter, we present  2D kinematic maps of the inner 250\,pc (radius) of NGC\,5929 and report the observation of double-component emission at the location of the radio hot spots, but also along a 50\,pc wide strip perpendicular to the radio jet using near-infrared integral field spectroscopy at Gemini North telescope. 

\section{Observations}

\object{NGC\,5929} was observed using the Gemini NIFS \citep{nifs03} operating with the adaptive optics module ALTAIR under the observation Programme GN-2011A-Q-43.  The observations covered the J and K$_{\rm l}$ spectral bands, resulting in a wavelength coverage from 1.14$\mu$m to 1.36$\mu$m  and 2.10$\mu$m to 2.54$\mu$m, respectively. The total on-source exposure time for each band was 6000\,s and the observations were split into 10 individual on source exposures plus 5 sky exposures for each band. The final data cubes cover the inner 3$^{\prime \prime}\times$3$^{\prime \prime}$ of NGC\,5929 at an angular resolution is 0\farcs12 (corresponding to $\sim$20\,pc at the galaxy)
and velocity resolution of $\sim$40\,km\,s$^{-1}$.

\section{Results} 

\begin{figure}
  \begin{tabular}{cc}
\multicolumn{2}{c}{\includegraphics[scale=0.32]{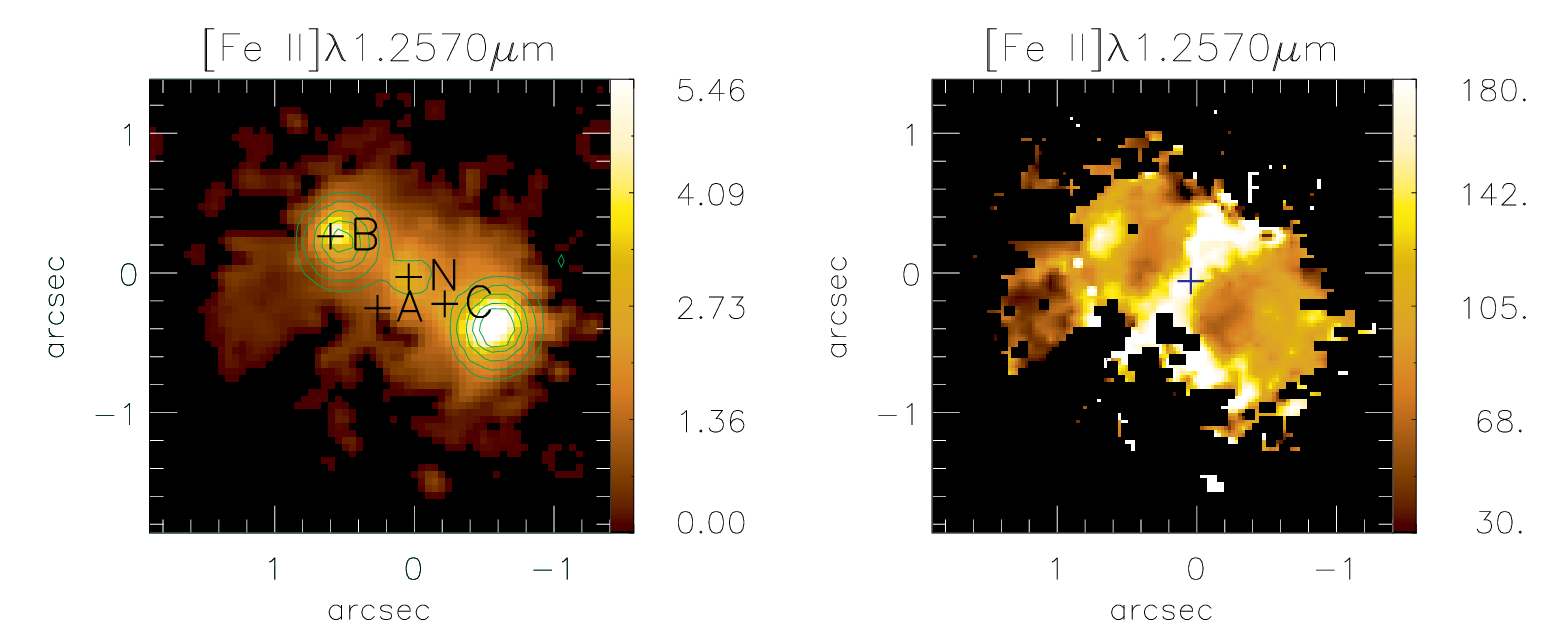}} \\
    \includegraphics[scale=0.17]{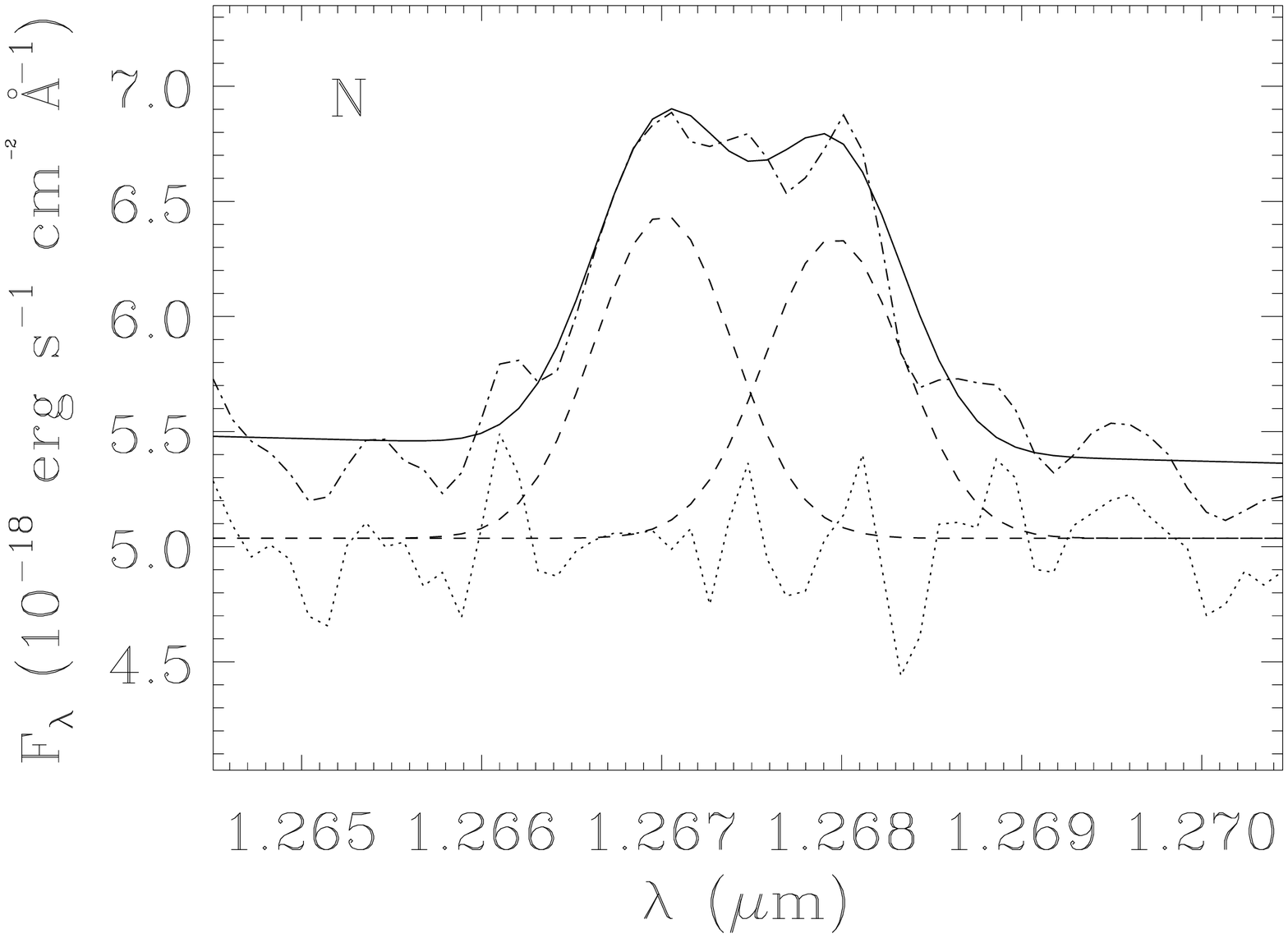}&
    ~\includegraphics[scale=0.17]{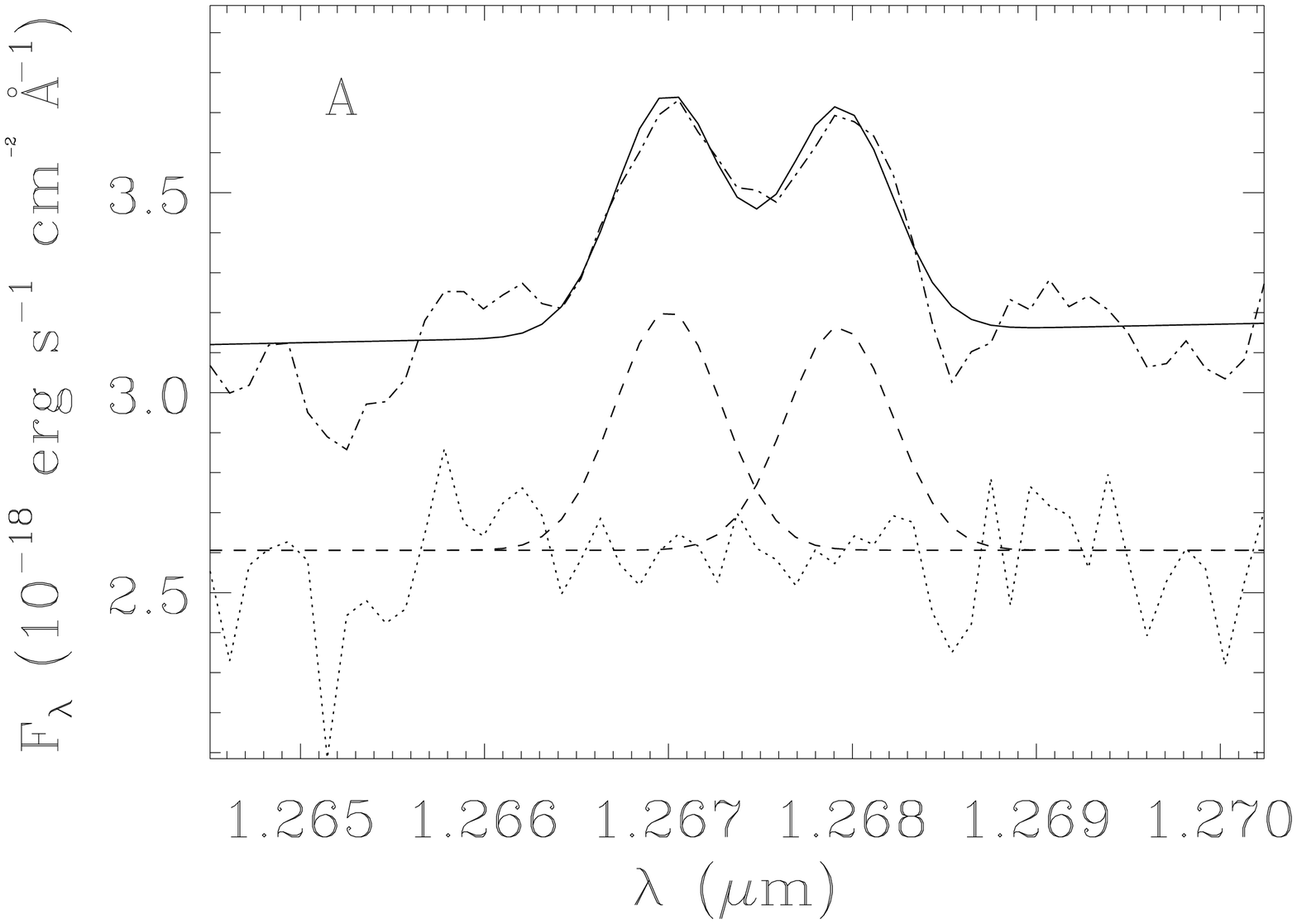}\\
    \includegraphics[scale=0.17]{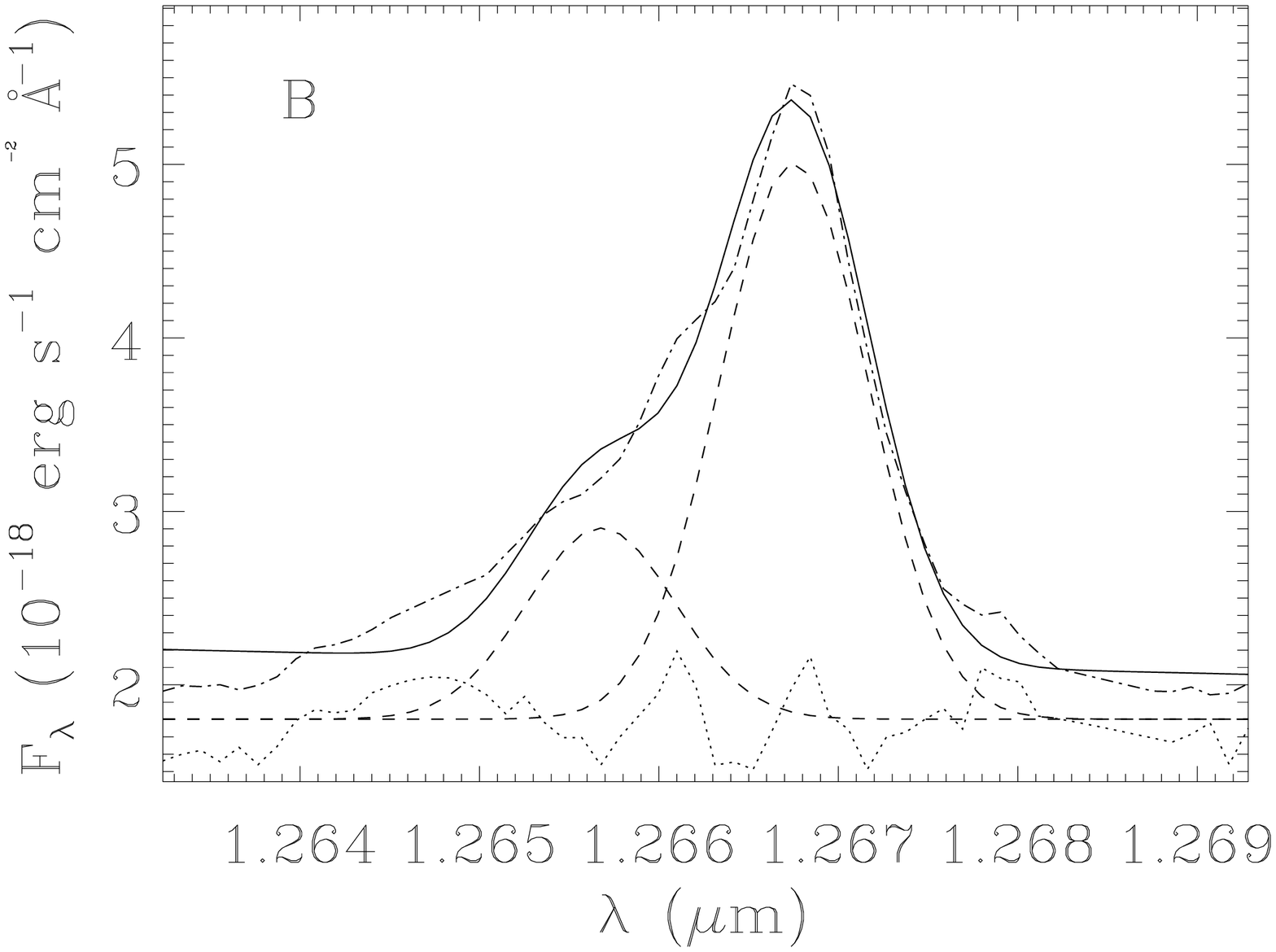}&
    ~\includegraphics[scale=0.17]{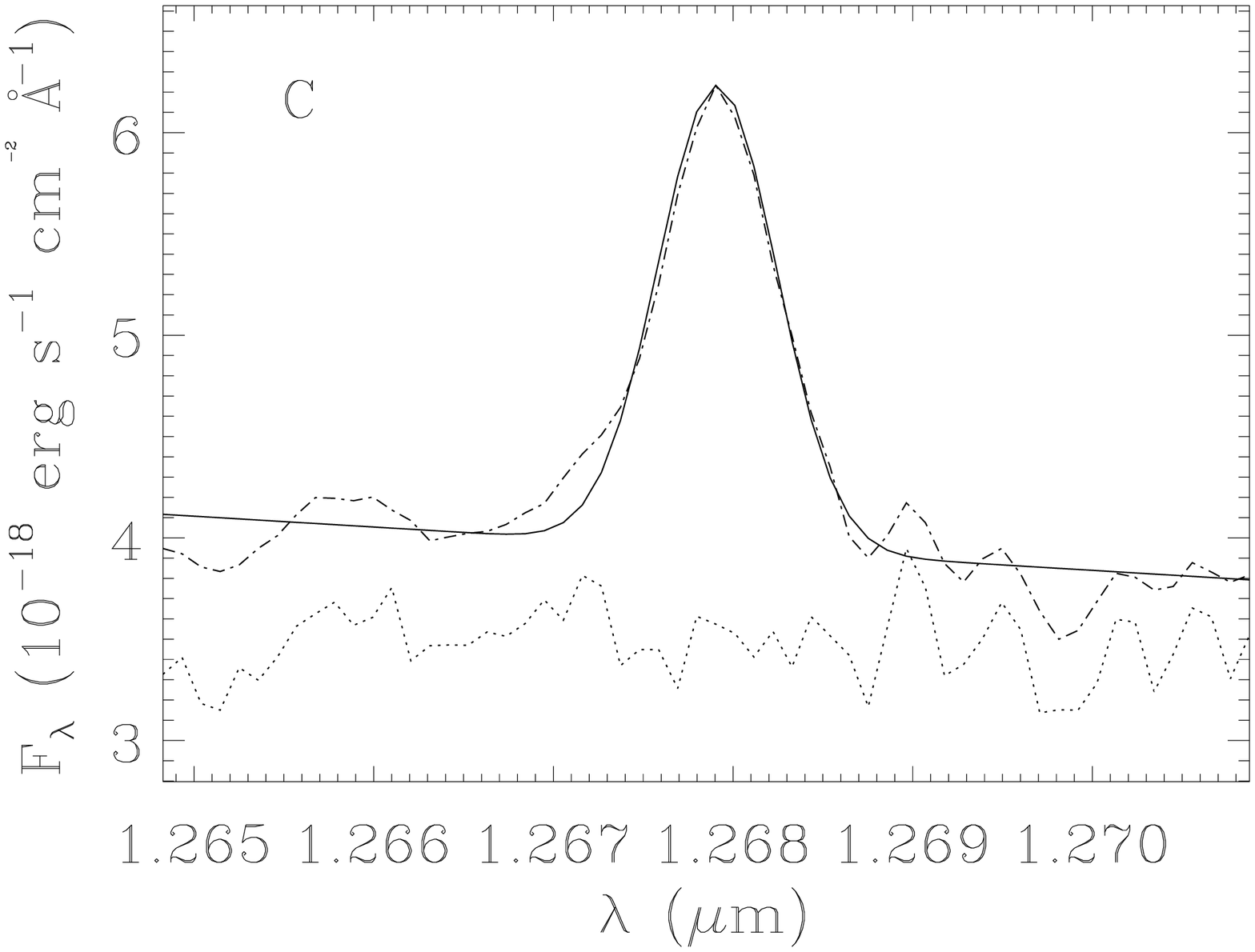}\\
  \end{tabular}
  \caption{Top-left: [Fe\,{\sc ii}]$\lambda1.25\mu$m flux distribution with contours of the radio image overlaid \citep{ulvestad89}. Top-right: $\sigma$ of the one component fits to the lines, where the SE-NW strip can be seen at PA$=-30^\circ$. The remaining panels show examples of the fitting of the [Fe\,{\sc ii}]$\lambda1.25\mu$m emission-line profile by two Gaussian curves (positions N, A and B) and by a single Gaussian (position C). The spectra are shown as dashed-dotted lines, the resulting fit as solid lines, each Gaussian component (plus a constant) as dashed lines and the residuals of the fit (plus a constant) as dotted lines.}
  \label{fits}
\end{figure}

In order to construct flux, velocity dispersion ($\sigma$) and centroid velocity maps for the emitting gas (\pii, \feii, \h2\ and H recombination lines), we first fitted the emission-line profiles by a single Gaussian, using the {\sc profit} fitting routine \citep{profit}. 
Fig.~\ref{fits} shows, in the top panels, the \feii$\lambda1.2570\,\mu$m  flux map in the left panel and the $\sigma$ map in the right panel.

The \feii\ velocity field obtained from the one-component fit is very similar to that of a previous study by \citet{oasis} for optical emission lines and we thus do not show it here. It presents the highest blueshifts of $\sim-$250\,\kms\ to  the north-east of the nucleus, while similar redshifts are seen to the south-west of it. The maximum velocity gradient is observed along PA$\approx$55$^\circ$. Our data has allowed also to obtain the stellar velocity field (Riffel et al., in preparation); we found it to be similar to the one obtained in the optical \citep{oasis}. It presents an amplitude of less than 100\,\kms, with the line of nodes oriented along PA$\approx$35$^\circ$. But the redshifts and blueshifts are observed to the north-east and south-west, respectively, thus contrary to what is observed in the gas velocity field, indicating that the gas is in counter rotation relative to the stars. 

The $\sigma$ map obtained from the one-component fit (top right panel of Fig.\ref{fits}) shows that the width of the Gaussians in a strip oriented along position angle PA=$-30^\circ$ (and also in a few other locations close to the radio hotspots) is almost twice the width of the profiles at other locations. A close inspection of these profiles reveal that at the locations where the profile is broader, they are in fact better fitted by two components and we repeated the fit using two components where necessary. The distinct profiles observed over the nuclear region and their fits are illustrated in the bottom panels of Fig.~\ref{fits}. We find that, besides the positions N and A, shown in Fig.~\ref{fits}, all profiles along PA=$-30^\circ$, in a strip $\approx$50\,pc wide, crossing the nucleus perpendicularly to the radio jet, have two components. We hereafter call this region SE-NW strip. Two components are also observed at the locations of the radio hotspots (e.g. position B), while for the remaining regions the profile is well reproduced by a single Gaussian (e.g. position C) with similar width to that of the individual components of the two-component profiles.

We now discuss the velocity fields derived for the different gas components. We found that the $\sigma$ of the gaussians fitted to the regions where only one component  is necessary is similar to that of each of the two components (where clearly two-components were necessary), being in the range 70--100\,\kms. We fitted two Gaussians to the \feii\ profile only in locations where the fit of one Gaussian resulted in $\sigma$ higher than 100 \kms. In regions where the profiles have $\sigma<$100\,\kms\ one Gaussian reproduces well the observed profile (see panel C of Fig.~\ref{fits}). Besides this criterium, if the two components present differences in velocity less than our spectral resolution ($\approx$40\,\kms) and/or one component had a flux lower than 5\% of the other one, we used the result of the fit of a single Gaussian. In addition, we kept the width of the two components the same (as this seems to work for the higher signal-to-noise regions), in order to have more robust fits, with fewer free parameters. 

\begin{figure*}
\begin{center}
\includegraphics[scale=0.65]{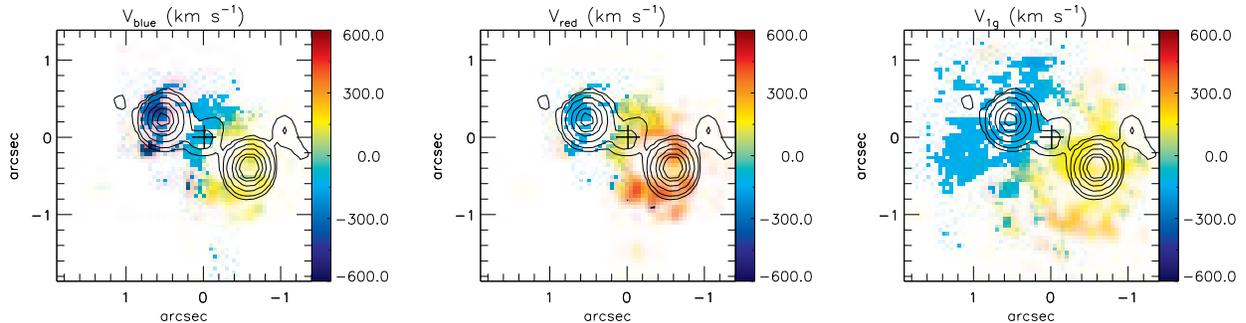}
\end{center}
    \caption{Centroid velocity maps obtained from the fit of the \feii\ emission line profile. Left: blue component of the double component fits; center: red component of the double component fits; right: velocities from the one component fits. The velocities are shown after the subtraction of the systemic velocity of the galaxy and the color bars represent velocities in units of \kms. }
    \label{2comp}
\end{figure*}

The right panel of Fig.~\ref{2comp} shows the velocity field for the region where only one Gaussian was necessary to fit the \feii\ profile, while the left and central panels show respectively, the velocity field of the blue and red components of the two Gaussian fits. Besides the SE-NW strip, two components are clearly present also at and around the radio hotspots.  It can be seen that the highest blueshifts -- of up to $-$600\,\kms\ -- correspond to the bluest components of the two component fit in the region of the north-east hotspot, while the highest redshifts -- of up to 600\,\kms --  are the reddest components of the two component fit in the region of the south-east hotspot.
 

\section{Discussion}

Optical Integral Field Spectroscopy (IFS) was obtained for the first time for NGC\,5929 by \citet{ferruit97} at an angular resolution of $\approx$1 arcsec and spectral resolution of R$\sim$1400. They point out the fact that the gas to the NE is observed in blueshift and to the SW in redshift, focusing on the apparent association between the gas emission and kinematics with the radio jet. They report the finding that the width of the [O{\sc iii}]$\lambda$5007 emission-line at the nucleus is twice its width to the NE and SW, attributing this to the interaction with the radio jet, but their measurements did not allowed to resolve the extent of the region at the nucleus with high velocity dispersions, neither conclude that the emission lines actually had two velocity components.


\citet{rosario10} found kinematic shock signatures attributed to the interaction of the radio jet with the NLR gas in NGC\,5929 using HST optical images, long-slit spectra and a 5\,GHz radio image. They compared the emission line profiles of H$\beta$ with those of [O\,{\sc iii}]$\lambda5007$ at distinct positions in the galaxy and found that they have similar widths at most locations, except in a region surrounding the north-eastern radio hot spot. There, the H$\beta$ profile is broader than that of [O\,{\sc iii}], suggesting that the low-ionization gas at the interaction site is more disturbed than the high-ionization gas. The authors attributed this broadening to a local enhancement of shock ionization due to the influence of the jet. The spectra shown by \citet{rosario10} does not include observations in the direction perpendicular to the radio jet, thus missing the SE-NW strip region, nor a nuclear spectrum, and only show the emission line profiles at four distinct locations.   

We can compare our results with those of \citet{rosario10} for the regions in common. Although the width of the \pb\ profile around  the position of the north-eastern radio hot spot is similar to those observed for H$_\beta$, the \pb\  profile does not present any hint of broadening or an additional component at this location relative to the surroundings. One possibility is that a second component is not detected at \pb\  because it is not bright enough, even though it can be detected at H$_\beta$, because H$_\beta$ is
6.2 times stronger than \pb\ \citep{osterbrock}. On the other hand, we observe two-components in the \feii\ profile at the locations of the radio hot spots. The \feii\ traces even lower ionization gas than \pb\, as the ion Fe\,{\sc ii} is formed in partially ionized regions, having an ionization potential of only 7.9\,eV  \citep{worden84}, while the \pb\ line originates from a fully ionized region \citep[e.g.][]{n1068-exc}.  Although many of our results in the near-IR support those by \citet{rosario10}, a new result we have obtained relative  to their study is the observation of the double-component profiles along the SE-NW strip, extending by $\approx$\,300\,pc and running perpendicularly to the main radio jet. This structure was not observed by \citet{rosario10} due to their restricted spatial coverage. 

\citet{whittle86} have suggested that the double components observed perpendicularly to the radio jet 
(which is approximately the orientation of the minor axis of the galaxy) 
could be interpreted as being originated in a rotating disk by the superposition of two components from opposite sides of the rotation curve which would have been unresolved by their observations. We note, however, that  even at our improved spatial resolution of $\approx$\,20\,pc, the emission-line profiles of the lines along the SE-NW strip are still double. Moreover, they are double not only along the presumed rotation major axis, where a velocity gradient is expected, but also along the presumed kinematic minor axis, where the velocity of a rotation field should be zero. Our data thus do not support the hypothesis the the two components observed along the minor axis of the galaxy are due to unresolved rotation.

\begin{figure}
\centering
\includegraphics[scale=0.45]{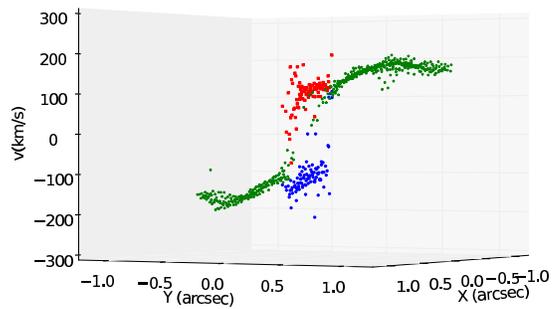}
\caption{Velocity 3D plot: The blue component is shown in blue, the red component in red and the single gaussian component in green.} 
\label{3d}
\end{figure}

In order to better inspect the gas kinematics close to the nucleus, we show in Fig.~\ref{3d}, a ``tri-dimensional rendering" of the velocity field: the spatial coordinates $x, y$ define the galaxy plane, and the centroid velocity is plotted as the third coordinate $z$. In this plot we show only the velocities corresponding to the two components along the SE-NW strip and those corresponding to the one-component fits (excluding the two components associated with the radio hotspots). The velocity field of the one component region (in green) seems to delineate a rotation curve, while the blueshifted and redshifted components (in blue and red, respectively) of the SE-NW strip seem to be detached form the one-component velocity field (even though the velocities are similar), indicating a distinct component.

\begin{figure*}
\begin{center}
\includegraphics[scale=0.45]{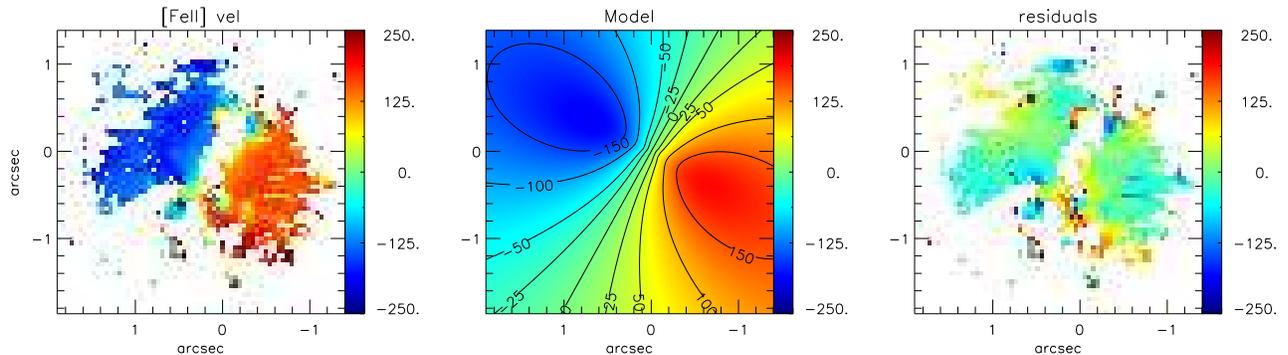}
\end{center}
    \caption{Left: \feii\ velocity field attributed to rotation; center: rotating disk model and right: residual map.}
    \label{model}
\end{figure*}

We have  fitted the one-component velocity field using a rotating disk model \citep[e.g.][]{mrk79}. The measured velocity field, the best fit rotation model and the residuals are shown in Fig.~\ref{model}. 
The best fit model shows quite a steep rotation curve,  
and presence of a few high residuals show that the velocity field is only approximately fitted by rotation.

As the gas velocity field rotates in the opposite direction to that of the stars, our interpretation is that this gas was captured in the interaction with NGC\,5930 and is now rotating around the nucleus of NGC\,5929 but still showing distortions from a ``regular" rotation field due to interaction with the companion. Inflow towards the center of this gas was probably the trigger of the nuclear activity in NGC\,5929. The onset of nuclear activity is associated with outflows originating in the accretion disk, and if a strong ``equatorial outflow" from the accretion disk could be generated, its interaction with the surrounding gas in the galaxy -- pushing it outwards -- could be the origin of the double component profiles seen along the SE-NW strip. The presence of such an outflow is nevertheless peculiar, as the best mapped outflows in nearby Seyfert galaxies are oriented approximately along the radio jet, not perpendicular to it \citep[e.g.][]{sb92,das05,das07,sb10}. 

 ``Equatorial" outflows from accretion disks have been recently proposed by \citet{li13}, being also compatible with a scenario in which the torus surrounding the AGN is actually a dusty wind emanating from the outer parts of the accretion disk \citep[e.g.][]{honig13,elitzur12,ivezic10,mor09,nenkova08,elitzur06}. We conclude that the peculiar velocity field observed in the SE-NW strip is thus probably the result of the interaction of an accretion disk equatorial outflow or torus wind with the surrounding gas in the galaxy. 

In Riffel et al. (in preparation) we find that the line ratios suggest the contribution of shocks to the excitation of the \feii\ and \h2\ lines along PA$=-30^\circ$, supporting the presence of an outflow there. In fact, \citet{su96} presented MERLIN and VLA radio images for NGC\,5929 at frequencies ranging from 15 to 0.4\,GHz and resolutions from 0\farcs05 to 0\farcs6. The radio emission at all frequencies is dominated by the three components seen in the contours of Fig.~\ref{fits}: the two hotspots plus a nuclear component. But the 0.4\,GHz image shows an additional extended emission oriented approximately perpendicular to the line connecting the two hotspots, along PA$\approx-30^\circ$ -- thus along the SE-NW strip. This extended radio emission has been attributed by \citet{su96}, to a flow of relativistic particles generated by the active nucleus flowing out perpendicularly to the radio jet. 
A similar scenario was proposed by \citet{pedlar93} for NGC\,4151. \citet{su96} discuss also an alternative scenario where the extended radio emission originates from starburst driven winds.  This scenario can nevertheless be discarded by our observations since the \feii/\pb\ and \h2/\bg\ ratios are higher than those expected for Starbursts  (Riffel et al., in preparation).

Finally, we have also found double components in the emission-line profiles from regions surrounding the radio hotspots. We note that one of the components has approximately the velocity of the one-component gas in the surrounding regions, while the other component shows a higher velocity, of up to $\approx$\,600\,\kms\ around the hotspots, in blueshift to the north-east and in redshift to the south-west. This indicates that the north-east side of the radio jet is tilted toward us. We support the interpretation of \citet{rosario10} that these high velocities are due to the radio jet interacting with the ambient medium, which in this case is the gas in counter rotation, pushing it away from the galaxy nucleus.

\section{Summary and Conclusions}

We used near-IR integral field spectroscopy of the inner 250~pc of NGC\,5929 at a spatial resolution of 20~pc and report the discovery of a 50\,pc wide strip crossing the nucleus perpendicularly to the radio jet (PA$=-30^\circ$, the SE-NW strip) and extending by 300~pc where the emission lines show two velocity components, one at $\approx-$150\,\kms\ and the other at $\approx$+150\,\kms.

We also show that the extended gas emission perpendicular to the SE-NW strip shows only one component, observed in blueshift to the north-east and redshift to the south-west, except around the region of the hotspots, where they are again double. This apparent rotation pattern is opposite to that of the stellar velocity field. We have fitted a rotation model to the one component gas and found a steep rotation curve, and residuals as high as 100\,\kms, evidencing deviations from simple rotation, which we attribute to the effect of the companion galaxy NGC\,5930. 

A possible scenario for the complex gas velocity field we have observed is the following: the interaction with NGC\,5930 has triggered the transfer of gas to the central region of NGC\,5929. The observation that the gas is in counter rotating relative to the stars supports an external origin. Part of this gas reaches the nucleus, where it triggers the nuclear activity, giving rise to the main radio jets which have the north-east side oriented towards us. The interaction of the radio jet with the rotating gas pushes the rotating gas to projected velocities along the line-of-sight of up to 600\,\kms. 

The nuclear activity seems also to have produced equatorial outflows along the accretion disk, whose interaction with the surrounding gas -- pushing it outwards -- originated the double components observed in the emission-line profiles along the
SE-NW strip. The presence of such outflows is also supported by the previous observation of faint radio emission perpendicular to the main radio jet. It is interesting to note also that other recent IFU studies of the central region of active galaxies by our group, such as in  \citet{couto13,schnorr13}, and in Lena et al. {\it in preparation} have also revealed broadened or double emission-line profiles perpendicular to the ionization axis, suggesting that outflows perpendicular to the main axis of the active nucleus may not be a rare phenomenon. These observations support recent models of equatorial accretion disk winds \citep{li13} and outflowing torus models \citep{honig13,elitzur12,ivezic10,mor09,nenkova08,elitzur06}.

\acknowledgments We thank the referee for valuable suggestions which helped to improve the present paper. This work is  based on observations obtained at the Gemini Observatory, which is operated by the Association of Universities for Research in Astronomy, Inc., under a cooperative agreement
with the NSF on behalf of the Gemini partnership: the National Science Foundation (United
States), the Science and Technology Facilities Council (United Kingdom), the
National Research Council (Canada), CONICYT (Chile), the Australian Research Council
(Australia), Minist\'erio da Ci\^encia e Tecnologia (Brazil) 
and Ministerio de Ciencia, Tecnologia e Innovaci\'on Productiva  (Argentina). This work has been partially supported by the Brazilian intitutions CNPq and FAPERGS.

{}   
\clearpage

\begin{thebibliography}{}


\bibitem[Cecil et al.(2002)]{cecil02}  Cecil, G. et al., 2002, ApJ, 568, 627.


\bibitem[Couto et al.(2013)]{couto13}  Couto, G. S., Storchi-Bergmann, T., Axon, D. J., Robinson, A., Kharb, P., Riffel, R. A., 2013, MNRAS, 435, 2982.


\bibitem[Das et al.(2005)]{das05}  Das, V. et al., 2005, AJ, 130, 945.


\bibitem[Das, Crenshaw \& Kraemer(2007)]{das07}  Das, V., Crenshaw, D. M., Kraemer, S. B., 2007, ApJ, 656, 699.

\bibitem[Elitzur \& Shlosman(2006)]{elitzur06} Elitzur M., Shlosman I., 2006, ApJ, 648, L10.

\bibitem[Elitzur(2012)]{elitzur12} Elitzur M., 2012, ApJL, 747, L33.

\bibitem[Falcke et al.(1998)]{falcke98}  Falcke, H., Wilson, A. S., \& Simpson, C., 1998, ApJ, 502, 199. 

\bibitem[Ferruit et al.(1997)]{ferruit97}  Ferruit, P.,  Pecontal, E., Wilson, A. S., Binette, L.Wilson, A. S., 1997, A\&A, 328, 493. 

\bibitem[Ferruit et al.(1999)]{ferruit99}  Ferruit, P.,  Wilson, A. S., Whittle, M., Simpson, C., Mulchaey, J. S., \& Ferland, G., 1999, ApJ, 523, 147. 

	
\bibitem[H\"onig et al.(2013)]{honig13} H\"onig, S. F. et al., 2013, MNRAS, 771, 87.


\bibitem[Ivezi\'c \& Elitzur(2010)]{ivezic10} Ivezi\'c, Z, Elitzur M., 2010, MNRAS, 404, 14151.


\bibitem[Keel(1985)]{keel85} Keel, W. C., 1985, Nature, 318, 43.

\bibitem[Li, Ostriker \& Sunyaev(2013)]{li13} Li, J., Ostriker, J., Sunyaev, R., 2013, ApJ, 767, 105L.



\bibitem[McGregor et al.(2003)]{nifs03}  McGregor, P. J. et al., SPIE, 4841, 1581, eds. Iye, M. \& Moorwood, A. F. M.

\bibitem[Mor, Netzer \& Elitzur(2009)]{mor09} Mor, R., Netzer, H., Elitzur M., 2009, ApJ, 705, 298.

\bibitem[Nenkova et al.(2008)]{nenkova08} Nenkova, M., Sirocky, M. Ivezi\'c, Z., Elitzur M., 2008, ApJ, 608, 147.

\bibitem[Osterbrock \& Ferland(2006)]{osterbrock} Osterbrock, D. E., Ferland, G. J., 2006, {\it Astrophysics of Gaseous nebulae and Active Galactic Nuclei}, University Science Books, Mill Valey, CA.


\bibitem[Pedlar et al.(1993)]{pedlar93} Pedlar, A., Kukula, M. J., Longley, D. P. T., Muxlow, T. W. B., Axon, D. J., Baum, S., O'Dea, C., Unger, S. W., 1993, MNRAS, 263, 471.

\bibitem[Riffel et al.(2006)]{eso428} Riffel, Rogemar A., Sorchi-Bergmann, T., Winge, C., Barbosa, F. K. B., 2006, MNRAS, 373, 2.



\bibitem[Riffel(2010)]{profit} Riffel, Rogemar A., 2010, Ap\&SS, 327, 239.

\bibitem[Riffel, Storchi-Bergmann \& Nagar(2010a)]{mrk1066-exc} Riffel, Rogemar A., Storchi-Bergmann, T. \& Nagar, N. M., 2010, MNRAS, 404, 166.


\bibitem[Riffel \& Storchi-Bergmann(2011a)]{mrk1066-kin} Riffel, Rogemar A. \& Storchi-Bergmann, T., 2011, MNRAS, 411, 469.



\bibitem[Riffel, Storchi-Bergmann \& Winge(2013)]{mrk79} Riffel, R. A., Storchi-Bergmann, T., Winge, C., 2013, 430, 2249.


\bibitem[Riffel et al.(2013a)]{n1068-exc} Riffel, Rogemar A., Storchi-Bergmann, Vale, T. B., McGregor, P. 2013, MNRAS, submited.






\bibitem[Rosario et al.(2010)]{rosario10} Rosario, D. J., Whittle, M., Nelson, C. H., \& Wilson, A. S., 2010, ApJ, 711, L94.

\bibitem[Schnorr M\"uller et al. (2013)]{schnorr13} Schnorr M\"uller, A, Storchi-Bergmann, T., Nagar, N. M., Robinson, A., Lena, D., Riffel, R. A., Couto, G. S., 2013, MNRAS, in press, http://arxiv.org/abs/1310.7916.

\bibitem[Stoklasov\'a et al.(2009)]{oasis} Stoklasov\'a, I., Ferruit, P., Emsellem, E., Jungwiert, B., P\'econtal, E., S\'anchez, S. F., 2009, A\&A, 500, 1287. 

\bibitem[Storchi-Bergmann et al. (1992)]{sb92} Storchi-Bergmann, T., Wilson, A. S. \& Baldwin, J. A. 1992, ApJ, 396, 45.


\bibitem[Storchi-Bergmann et al.(2010)]{sb10} Storchi-Bergmann, T., Sim\~oes Lopes, R., McGregor, P. Riffel, Rogemar A., Beck, T., Martini, P., 2010, MNRAS, 402, 819.



\bibitem[Su et al.(1996)]{su96}  Su, B. M., Muxlow, T. W. B., Pedlar, A. Holloway, A. J., Steffen, W., Kukula, M. J., \& Mutel, R., L., 1996, MNRAS, 279, 1111.

\bibitem[Ulvestad \& Wilson(1984)]{ulvestad84}  Ulvestad, J., S., \& Wilson, A. S., 1984, ApJ, 285, 439. 

\bibitem[Ulvestad \& Wilson(1989)]{ulvestad89}  Ulvestad, J., S., \& Wilson, A. S., 1989, ApJ, 343, 659. 

\bibitem[Wilson \& Keel(1989)]{wilson89}  Wilson, A. S., \& Keel, W. C.,  1989, AJ, 98, 1581. 

\bibitem[Whittle et al.(1986)]{whittle86} Whittle, M., Haniff, C. A., Ward, M. J., Meurs, E. J. A., Pedlar, A., Unger, S. W., Axon, D. J., Harrison, B. A., 1986, MNRAS, 222, 189. 


\bibitem[Whittle \& Wilson(2004)]{whittle04}  Whittle, M., \& Wilson, A. S., 2004, AJ, 127, 606. 


\bibitem[Worden et al.(1984)]{worden84} Worden, E.F., Comaskey, B., Densberger, J., Christensen, J., McAfee, J.M., Paisner, J.A., Conway, J.G. 1984, J. Opt. Soc. Am. B: Opt. Phys., 1:2, 314.






   
\end{thebibliography}
\end{document}